\documentclass[
preprint,
 amsmath,amssymb,
 aps,
showkeys,
showpacs 
]{revtex4}

\usepackage{comment}
\usepackage{graphicx}
\usepackage{dcolumn}
\usepackage{bm}

\usepackage{revsymb4-2}

\graphicspath{{figs/}}

\newcommand{\lsim}
{\;\raisebox{-.3em}{$\stackrel{\displaystyle <}{\sim}$}\;}

\newcommand\ReDiag{\mathop{%
  \raise .5pt\hbox{[}%
  \widetilde{\mathrm{Re}}%
  \raise .5pt\hbox{]}}}
\newcommand\ReOffDiag{\mathop{%
  \raise .5pt\hbox{$\llbracket$}%
  \widetilde{\mathrm{Re}}%
  \raise .5pt\hbox{$\rrbracket$}}}

\newcommand\Sl{\tilde l}

\newcommand\Slpm{\tilde l^\pm}

\newcommand\msl[1]{m_{\Sl_{#1}}}

\newcommand\ino[1]{\tilde\chi_{#1}}

\newcommand\chapm[1]{\ino{#1}^\pm}

\newcommand\cha{\chapm}

\newcommand\neu[1]{\ino{#1}^0}
\newcommand\mneu[1]{m_{\neu{#1}}}

\newcommand\citere[1]{Ref.~\cite{#1}}
\newcommand\citeres[1]{Refs.~\cite{#1}}

\newcommand{\CP}{{\cal CP}}
\newcommand{\cp}{{\CP}}

\newcommand{\tev}{\,\, \mathrm{TeV}}
\newcommand{\gev}{\,\, \mathrm{GeV}}

\newcommand\MO{\texttt{MicrOMEGAs}}
\newcommand\CM{\texttt{CheckMATE}}

\newcommand{\sig}{\sigma}

\def\reffi#1{\mbox{Fig.~\ref{#1}}}

\def\De{\Delta}

\def\gmin2{\ensuremath{(g-2)_\mu}}
\def\amu{\ensuremath{a_\mu}}

\newcommand{\ssi}{\ensuremath{\sig_p^{\rm SI}}}

\begin{document}

\preprint{IFT--UAM/CSIC--22-001}

\title{SUSY Dark Matter Direct Detection Prospects
based on \boldmath{\gmin2}}

\author{Manimala Chakraborti}
 \email{mani.chakraborti@gmail.com}
\affiliation{%
Astrocent, Nicolaus Copernicus Astronomical Center
of the Polish Academy of Sciences,
ul. Rektorska 4, 00-614 Warsaw, Poland
}%


\author{Sven Heinemeyer}
\email{Sven.Heinemeyer@cern.ch}
\thanks{(speaker)}
\affiliation{
IFT (UAM/CSIC),
Cantoblanco, 28049, Madrid, Spain
}%

\author{Ipsita Saha}
\email{ipsita.saha@ipmu.jp}
\affiliation{%
Kavli IPMU (WPI), UTIAS, University of Tokyo, Kashiwa,
Chiba 277-8583, Japan
}%


\date{\today}

\begin{abstract}
An electroweak (EW) sector of the Minimal Supersymmetric Standard Model
(MSSM) with masses of a few hundred GeV can account for variety of
experimental data, assuming the lightest neutralino to be the lightest
supersymmetric (SUSY) particle: the non-observation at the LHC, 
searches owing to their small production cross sections, the results for the
(upper limit of the)
Dark Matter (DM) relic abundance and the DM Direct Detection (DD) limits.
Such a light EW sector can in particular explain the reinforced
$4.2\,\sig$ discrepancy between the experimental result for 
\gmin2, and its Standard Model (SM) prediction.  
Using the improved limits on \gmin2, we review the predictions for the
future prospects of the DD experiments. This analysis is performed for
several different realizations of DM in the MSSM:
bino, bino/wino, wino and higgsino DM. 
We find that higgsino, wino and one type of bino scenario can be
covered by future DD experiments. Mixed bino/wino and another type of
bino DM can reach DD cross sections below the neutrino floor. In these 
cases future collider experiments must cover the remaining parameter space.
\end{abstract}

\maketitle



\section{Introduction}
\label{sec:intro}

Searches for Dark Matter (DM) is one of the main objectives in today's
particle and astroparticle physics. Searches at the LHC (or other
collider experiments) are complementary to the searches in ``direct
detection'' (DD) experiments. 
Among the Beyond the Standard Model (BSM) theories that predict a
viable DM particle the Minimal Supersymmetric Standard Model  
(MSSM)~\cite{Ni1984,Ba1988,HaK85,GuH86} is one of the leading candidates.
Supersymmetry (SUSY) predicts two scalar partners for all Standard
Model (SM) fermions as well as fermionic partners to all SM bosons. 
The MSSM requires two Higgs doublets, resulting
in five physical Higgs bosons:
the light and heavy $\cp$-even Higgs bosons, 
$h$ and $H$, the $\cp$-odd Higgs boson, $A$, and the charged Higgs bosons,
$H^\pm$.
The neutral SUSY partners of the (neutral) Higgs and electroweak (EW) gauge
bosons gives rise to the four neutralinos, $\neu{1,2,3,4}$.  The corresponding
charged SUSY partners are the charginos, $\cha{1,2}$.
The SUSY partners of the SM leptons and quarks are the scalar leptons
and quarks (sleptons, squarks), respectively.
The lightest SUSY particle (LSP) is naturally the lightest neutralino,
$\neu1$. It can make up the full DM content of the
universe~\cite{Go1983,ElHaNaOlSr1984}, or, depending on its nature
only a fraction of it. In the latter case, an additional DM component could
be, e.g., a SUSY axion~\cite{Bae:2013bva},
which would then bring the total DM density into agreement with the
experimental measurement.

In \citeres{CHS1,CHS2,CHS3,CHS4} we performed a comprehensive analysis of the EW
sector of the MSSM, taking into account all
relevant theoretical and experimental constraints. 
The experimental results comprised the direct searches at the
LHC~\cite{ATLAS-SUSY,CMS-SUSY}, the DM relic abundance~\cite{Planck}
the DM direct detection (DD) experiments~\cite{XENON,LUX,PANDAX} and in
particular the  deviation of the anomalous magnetic moment
of the muon.
%
Five different scenarios were analyzed, classified by the mechanism
that brings the LSP relic density into agreement with the measured
values. The scenarios differ by the Next-to-LSP (NLSP), or
equivalently by the mass hierarchies between the mass scales
determining the neutralino, chargino and slepton masses.
These mass scales are the gaugino soft-SUSY breaking parameters $M_1$
and $M_2$, the Higgs mixing parameter $\mu$ and the slepton soft
SUSY-breaking parameters $\msl{L}$ and $\msl{R}$, see
\citeres{CHS1,CHS2,CHS3,CHS4} for a detailed description. 
The five scenarios can be summarized as follows, 
\begin{itemize}
\item[(i)]
higgsino DM ($\mu < M_1, M_2, \msl{L}, \msl{R}$),
DM relic density is only an upper bound (the rull relic density implies
$\mneu1 \sim 1 \tev$ and \gmin2\ cannot be fulfilled), 
$m_{\rm (N)LSP} \lsim 500 \gev$ with $m_{\rm NLSP} - m_{\rm LSP} \sim 5 \gev$;
\item[(ii)]
wino DM ($M_2 < M_1, \mu, \msl{L}, \msl{R}$),
DM relic density is only an upper bound, (the rull relic density implies
$\mneu1 \sim 3 \tev$ and \gmin2\ cannot be fulfilled),
$m_{\rm (N)LSP} \lsim 600 \gev$ with $m_{\rm NLSP} - m_{\rm LSP} \sim 0.3 \gev$;
\item[(iii)]
bino/wino DM with $\cha1$-coannihilation ($M_1 \lsim M_2$),
DM relic density can be fulfilled, $m_{\rm (N)LSP} \lsim 650\, (700) \gev$;
\item[(iv)]
bino DM with $\Slpm$-coannihilation case-L ($M_1 \lsim \msl{L}$),
DM relic density can be fulfilled, $m_{\rm (N)LSP} \lsim 650\, (700) \gev$;
\item[(v)]
bino DM with $\Slpm$-coannihilation case-R ($M_1 \lsim \msl{R}$),
DM relic density can be fulfilled, $m_{\rm (N)LSP} \lsim 650\, (700) \gev$.
\end{itemize}

Recently the ``MUON G-2'' collaboration published the results of their
Run~1 data~\cite{Abi:2021gix}, which is within $0.8\,\sig$ in
agreement with  the older BNL result on \gmin2~\cite{Bennett:2006fi}.
The combined measurement yields a deviation from the SM prediction of
$\De\amu = (25.1 \pm 5.9) \times 10^{-10}$, corresponding to $4.2\,\sig$.
Imposing this limit on the MSSM parameter space allows to set {\it upper} 
limits on the EW sector. A list of works that explain the \gmin2\ result
with in SUSY can be found in \citeres{CHS4}. 

Here we review the results in the five scenarios as obtained in
\citere{CHS4} for the expectations of the future DD experiments.
We take into account the projections for
the exclusion reach of XENON-nT~\cite{Aprile:2020vtw} and of the LZ
experiment~\cite{LZ} (which effectively agree with each other).
We also include the projections of the 
DarkSide~\cite{DarkSide} and Argo~\cite{Argo} experiments,
which can go down to even lower cross sections, 
as well as the neutrino floor (NF)~\cite{neutrinofloor}.


\section {Relevant constraints}
\label{sec:constraints}

The SM prediction of \amu\ is given by~\cite{Aoyama:2020ynm}.
The comparison with the combined experimental new world average, based
on \citeres{Abi:2021gix,Bennett:2006fi} yields a deviation of 
$\De\amu = (25.1 \pm 5.9) \times 10^{-10}$, corresponding to $4.2\,\sig$.
The prediction of \gmin2\ in the MSSM is calculated using
{\tt GM2Calc}~\cite{Athron:2015rva}, implementing two-loop corrections
from \cite{vonWeitershausen:2010zr,Fargnoli:2013zia,Bach:2015doa}
(see also \cite{Heinemeyer:2003dq,Heinemeyer:2004yq}).
Vacuum stability constraints are taken into account with
the public code {\tt Evade}~\cite{Hollik:2018wrr,Ferreira:2019iqb}.
All relevant SUSY searches for EW particles are taken into account,
mostly via \CM~\cite{Drees:2013wra,Kim:2015wza,Dercks:2016npn} (see 
\citere{CHS1} for details on many analyses newly implemented by our group).
For the DM relic density constraints we use the latest result from
Planck~\cite{Planck}, 
either as a direct measurement, or as an upper bound.
The relic density in the MSSM is evaluated with
\MO~\cite{Belanger:2001fz,Belanger:2006is,Belanger:2007zz,Belanger:2013oya}.
For the DD DM constraints, we use the results for the spin-independent
DM scattering cross-section $\ssi$ from
XENON-1T~\cite{XENON} experiment.
The theoretical predictions are evaluated using the public code
\MO.
Details about the parameter scan performed in the five scenarios can be
found in \citeres{CHS1,CHS2,CHS3,CHS4}. 


\section{Results}
\label{sec:results}

\begin{figure}[htb!]
\centering
\includegraphics[width=0.45\textwidth]{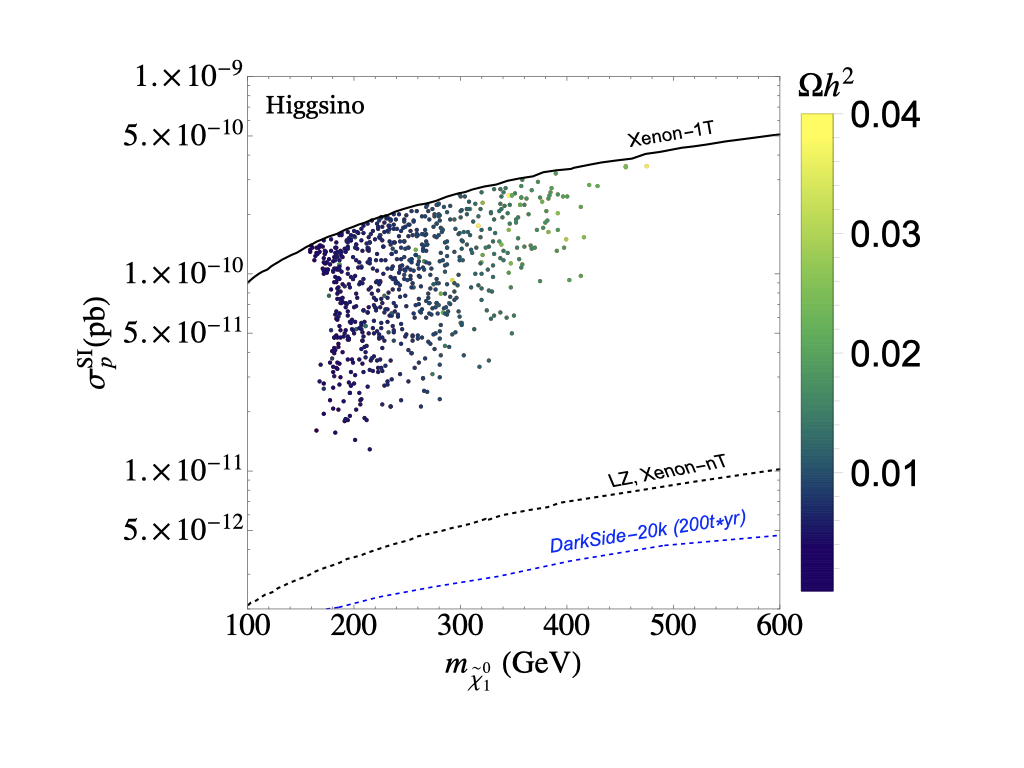}
\includegraphics[width=0.45\textwidth]{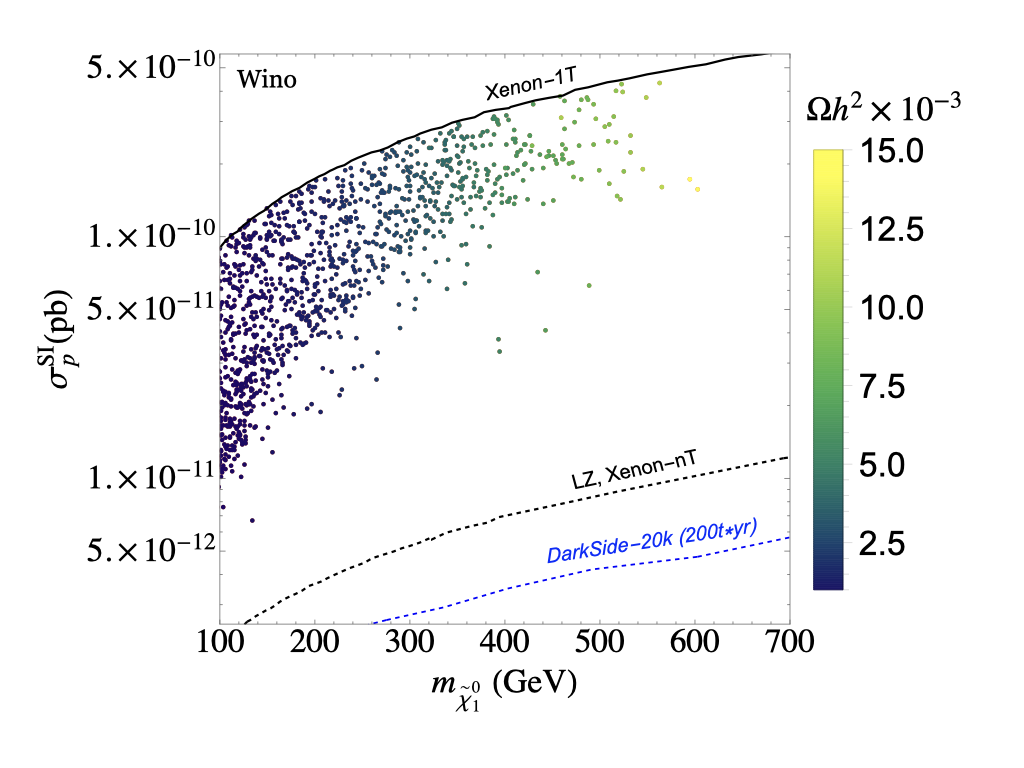}\\
\includegraphics[width=0.45\textwidth]{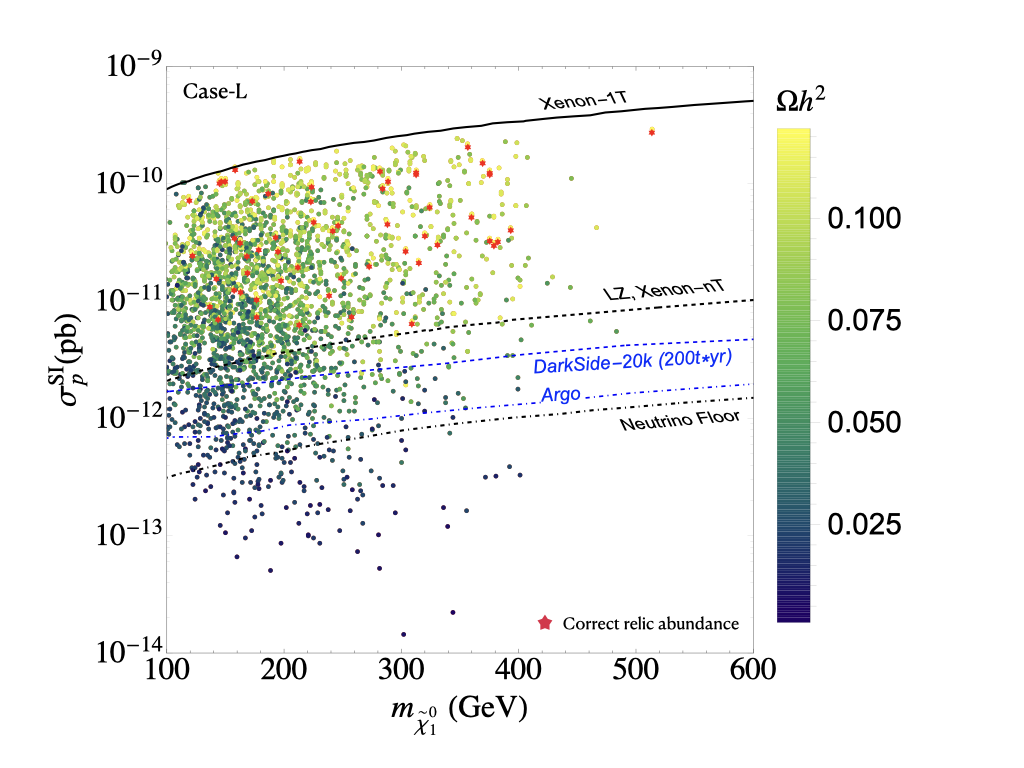}
\includegraphics[width=0.45\textwidth]{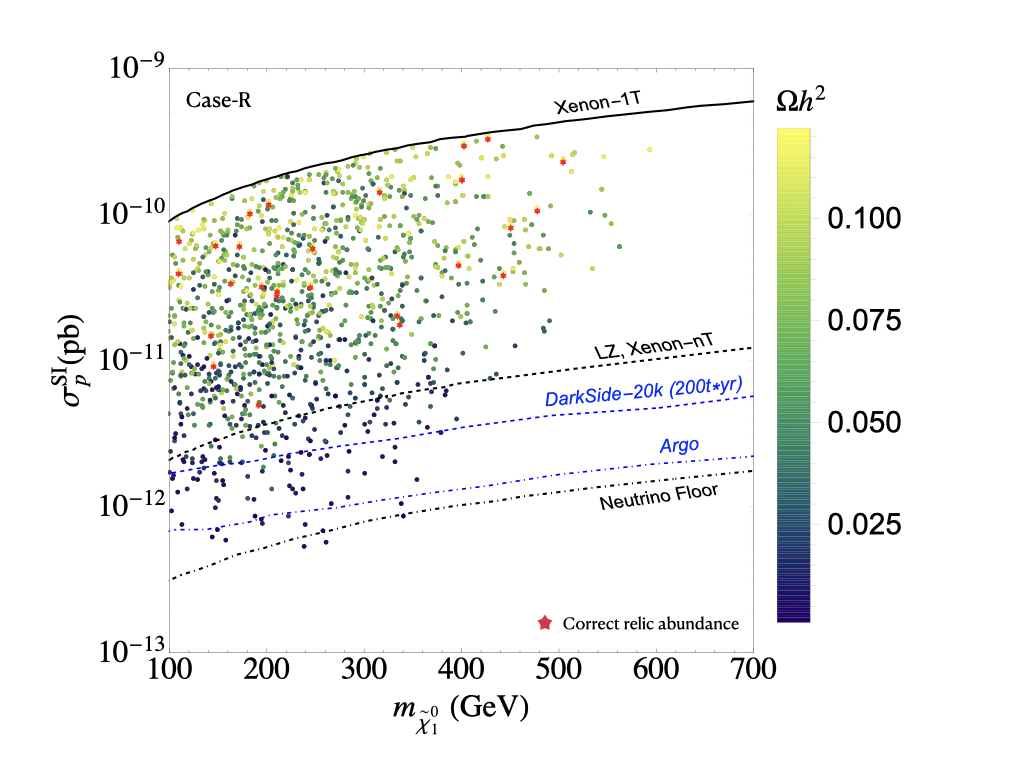}\\
\includegraphics[width=0.45\textwidth]{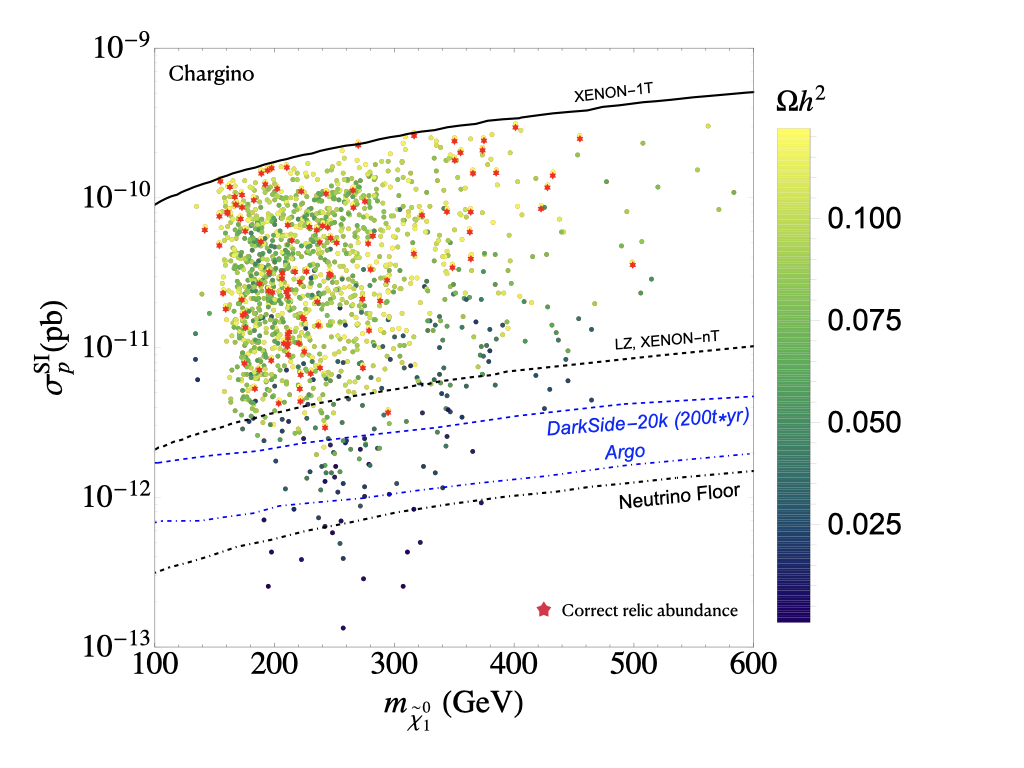}
\caption{The results of our parameter scan in the five DM scenarios
in the  $\mneu1$--$\ssi$ plane.
The color code indicates the DM relic density.
Red points are in full agreement with the Planck measurement.
}
\label{fig:results}
\end{figure}

In this section we review our results for the DM DD prospects in the
five scenarios~\cite{CHS4}. They are summarized in \reffi{fig:results},
where we show the $\mneu1$--$\ssi$ planes for higgsino DM (upper left),
wino DM (upper right), bino DM case-L (middle left) and case-R (middle
right) and bino/wino DM with $\cha1$-coannihilation (lower plot).
The color code indicates the DM relic density, where the 
red points are in full agreement with the Planck measurement.
The black dashed, blue dashed, blue dot-dashed and black dot-dashed
lines indicate the prospects for LZ/Nenon-nT, DarkSide, Argo and the
NF, respectively. 

One can observe that in for higgsino and wino DM all points will
be covered by the next round of DD experiments, LZ and/or Xenon-nT
are sufficient to cover the whole parameter space. The situation is
different for bino DM case-L/R and bino/wino DM. In these three cases
cases a large part of the allowed points cannot be probed by
LZ/Xenon-nT. The Argon-based experiments can cover a substantially
larger part of the allowed parameter space, where he parameter points
giving the full DM relic density are mostly covered by LZ/Xenon-nT, but
DarkSide/Argo might be needed in the case of $\cha1$-coannihilation, as
can be seen in the lower plot.

However, in all three scenarios some parameter points are allowed even
below the NF, which makes them unaccessible to current DD
techniques (see \citere{CHS4} for a short discussion on future
directional detection techniques). The allowed parameter spaces below
the NF are relatively restricted in the LSP mass, which is
bound to be $\mneu1 \lsim 400 \gev$. This makes them relatively easily
accessible to possible future $e^+e^-$ colliders. As was demonstraded in
\citere{CHS4}, all points below the NF can indeed be covered
by an $e^+e^-$ collider with $\sqrt{s} \lsim 1 \tev$. On the other hand,
at the HL-LHC these points may still remain elusive due to the small
mass splitting between the NLSP and the LSP.\\[.3em]
{\bf Acknowledgements:} S.H.\ thanks the organizers of the Lomonosov
conference for the invitation to present this work -- and hopes for an
in-person conference next time. :-)


\newcommand\jnl[1]{\textit{\frenchspacing #1}}
\newcommand\vol[1]{\textbf{#1}}


\end{document}